\documentclass[preprint,12pt]{elsarticle}

\usepackage{amssymb}

\usepackage{color}
\usepackage{lineno}
\usepackage{url}
\usepackage{ulem}
\usepackage{lineno}

\journal{Journal of Molecular Liquids}

\begin{document}

\begin{frontmatter}

%% Title, authors and addresses

%% use the tnoteref command within \title for footnotes;
%% use the tnotetext command for theassociated footnote;
%% use the fnref command within \author or \address for footnotes;
%% use the fntext command for theassociated footnote;
%% use the corref command within \author for corresponding author footnotes;
%% use the cortext command for theassociated footnote;
%% use the ead command for the email address,
%% and the form \ead[url] for the home page:
%% \title{Title\tnoteref{label1}}
%% \tnotetext[label1]{}
%% \author{Name\corref{cor1}\fnref{label2}}
%% \ead{email address}
%% \ead[url]{home page}
%% \fntext[label2]{}
%% \cortext[cor1]{}
%% \address{Address\fnref{label3}}
%% \fntext[label3]{}

\title{Improved data analysis for molecular dynamics in liquid CCl$_4$}

%% use optional labels to link authors explicitly to addresses:
%% \author[label1,label2]{}
%% \address[label1]{}
%% \address[label2]{}

\author[IINa]{Shinya~Hosokawa\corref{mycorrespondingauthor}}
\cortext[mycorrespondingauthor]{Corresponding author}
\ead{shhosokawa@kumamoto-u.ac.jp}
\author[FU]{Koji~Yoshida}

\address[IINa]{Institute of Industrial Nanomaterials, Kumamoto University, Kumamoto 860-8555, Japan}
\address[FU]{Department of Chemistry, Faculty of Science, Fukuoka University, Fukuoka 814-0180, Japan}

\begin{abstract}
%% Text of abstract
Previously reported inelastic x-ray scattering spectra of a typical van der Waals molecular liquid CCl$_4$ were reanalyzed by using a generalized Langevin formalism with a memory function including a thermal and two viscoelastic relaxation processes together with a simple sparse modeling. The obtained excitations of longitudinal acoustic phonons show a largely positive deviation from the hydrodynamic value by about 57\%, much larger than about 37\% by the previously reported damped harmonic oscillator result. Such large values of fast sounds in molecular liquids being larger than about 15-20\% of typical liquid metals are interpreted as extra energy-losses for terahertz phonons by vibrational and rotational motions of molecules. The rates of the fast and slow viscoelastic relaxations in the memory function at low $Q$ indicate large values, about 0.5 and 2 ps, which correspond to the vibrational and rotational motions of CCl$_4$ molecules, respectively. These values are larger than those of the typical polar molecular liquid acetone, which may reflect the heavier atomic mass of CCl$_4$. The $Q$ dependences of the viscoelastic relaxation rates are discussed in terms of lifetime and propagating length of the terahertz phonon oscillations. The microscopic kinematic longitudinal viscosity rapidly decreases with $Q$ from a reasonable macroscopic value.   
\end{abstract}

%%Graphical abstract
\begin{graphicalabstract}

\includegraphics[width=80mm]{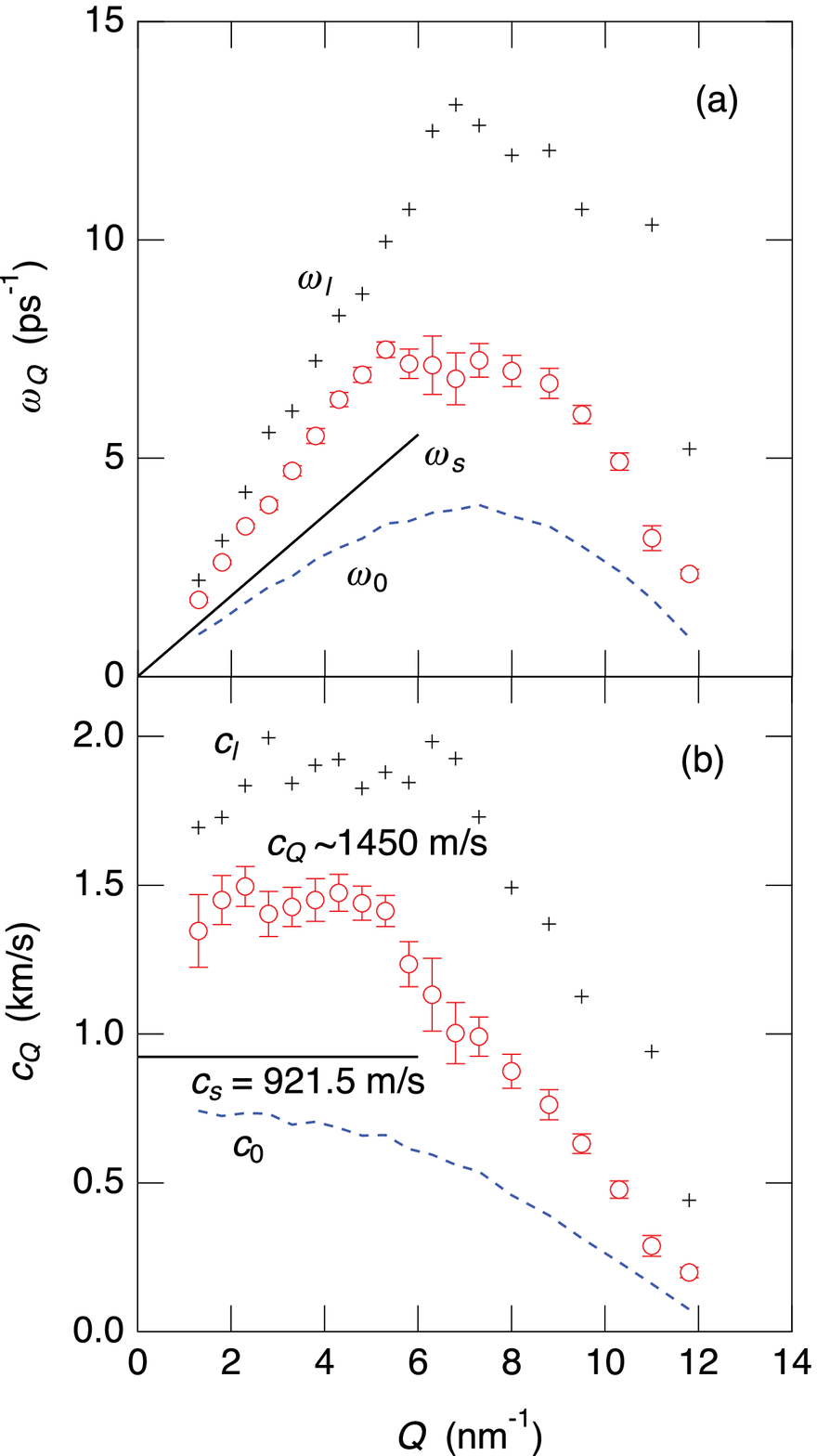}

\end{graphicalabstract}

%%Research highlights
\begin{highlights}
\item Generalized  Langevin formalism was applied to inelastic x-ray scattering spectra of liquid CCl$_4$
\item A large deviation from hydrodynamic prediction was found in phonon dispersion relation
\item Fast sound in terahertz oscillations originates from viscoelastic damping by vibrational and rotational molecular motions
\end{highlights}

\begin{keyword}
Van der Waals liquid \sep Density fluctuation \sep Inelastic x-ray scattering \sep Viscoelastic relaxation \sep Generalized Langevin formalism \sep Sparse modeling
%% keywords here, in the form: keyword \sep keyword

%% PACS codes here, in the form: \PACS code \sep code

%% MSC codes here, in the form: \MSC code \sep code
%% or \MSC[2008] code \sep code (2000 is the default)

\end{keyword}

\end{frontmatter}

%% \linenumbers

%% main text
\section{Introduction}
Liquid CCl$_4$ is a typical and simple molecular liquid, having a symmetric molecular structure of Cl tetrahedra around a central C atom. The molecules correlate with each other by a weak van der Waals force. Thus, liquid CCl$_4$ has been considered as an analogue to liquid rare gases, such as liquid Ar \cite{ArND}. Nevertheless, distinct differences in the structure and dynamics are realized between the single particle of rare-earth elements and molecules made of the atomic group. The static structure of liquid CCl$_4$ was investigated by a neutron diffraction (ND) experiment \cite{Misawa}, which revealed that beyond the atomic configurations of a CCl$_4$ single molecule, orientations between neighboring molecules should be taken into account in addition to a dense-packed scheme of molecules. Such an orientational correlation appeared in the temperature dependence of the structure factors, $S(Q)$, in the $Q$ range from 20 to 35 nm$^{-1}$. An edge-to-face orientation engagement is suggested to be preferred, i.e., a Cl atom in a neighboring molecule is nested into a hollow of the centering molecule, and one of the Cl atoms in the central molecule is also located near a hollow of the neighboring molecule. The molecular orientation was also investigated in combination with ND and reverse Monte Carlo (RMC) modeling by Pusztai and McGreevy \cite{Pusztai},  as wel as that with x-ray diffraction (XRD), ND, and RMC by Pothoczki et al. \cite{Pothoczki}, and a different intermolecular orientation of an edge-to-edge arrangement was suggested to be preferred. 

The transverse and rotational molecular motions in liquid CCl$_4$ were studied by high resolution (15-50 $\mu$eV) inelastic neutron scattering (INS) in combination with a molecular dynamics (MD) simulation \cite{Bermejo, Chahid, Garcia-Hernandez}. The experimental and calculated spectra were analyzed by simple analytical models. Due to the kinetic restrictions of the INS technique \cite{Burkel}, however, high-energy neutrons of 37.0 meV were used for the experiment, resulting in a rather poor energy resolution of 3.3 meV full-width at half-maximum (FWHM). In particular, the INS results concerning the phonon excitations were limited to their lower and higher limits, and the detailed discussion largely relied on the MD simulations.

In 2004, we carried out inelastic x-ray scattering (IXS) experiments on liquid CCl$_4$ to investigate the longitudinal acoustic (LA) phonon excitations more precisely \cite{KamiyamaCCl4JPSJ}. We used the best IXS spectrometer installed at BL35XU of the SPring-8 \cite{BaronJPCS} at that time, having a good energy resolution of about 1.5 meV FWHM and no kinematic restrictions. The observed IXS spectra demonstrate the existence of the LA modes at small $Q$ values, although the excitation peaks are highly damped. The spectra were analyzed by a damped harmonic oscillator (DHO) model \cite{Fak}, and the dispersion relation of the LA  modes shows a positive deviation from the hydrodynamic value by about 37\%. This value is much larger than those of liquid metals of about 15-20\% \cite{ScopignoRev}. 

The positive deviations of sound velocity of LA phonons for liquid Rb and Cs were obtained by INS, and interpreted by the framework of generalized hydrodynamics containing a solid-like shear elasticity on the ps time scale \cite{MorkelRbCs}. Positive dispersions in other liquid metals can be, in principle, explained by the same features of dynamical mechanism. However, a larger magnitude of such a {\it fast} sound was not observed so far, except in liquid Te \cite{HosokawaTe, KajiharaTe}, which was not considered as a densely packed liquid metal. Therefore, extra energy losses should be included to understand such a large magnitude of the fast sound in a molecular liquid in the terahertz frequency region. Our previous analytical method of the DHO model, however, gave only the phonon excitation energies and widths, and was impossible for the further analysis of the dynamic properties in liquid CCl$_4$. 

Very recently, we measured IXS on another molecular liquid, liquid acetone, to investigate the dynamics of this polar molecular liquid \cite{HosokawaAcetone}. For understanding the microscopic dynamics in detail, we analyzed the experimental IXS data by using a generalized Langevin formalism (GLF) with a memory function, which provides a detailed mechanism of particle dynamics, such as double viscoelastic relaxation rates and a microscopic viscosity. From the GLF analysis, we found that 1) the dispersion relation of the LA modes shows a highly positive deviation by about 65\%, 2) the fast and slow viscoelastic relaxations in the memory function correspond to the vibrational dynamics and reorientation correlations of the acetone molecules, and 3) the longitudinal kinetic viscosity rapidly decreases with $Q$ from a reasonable hydrodynamic value. 

In this article, we recall the van der Waals molecular liquid CCl$_4$, and reanalyze and clarify if the particle dynamics obtained from the GLF is surely related to the macroscopic dynamical data, and clarify the origin of the large positive deviation of the sound velocity. We again apply a simple sparse modeling to the analysis so that  the obtained dynamical parameters for the IXS spectra smoothly change with varying $Q$. In the article, the analytical procedures are given in Sec. 2, and the results are presented in Sec. 3. We discuss the dispersion relation of the LA phonon excitation, the viscoelastic relaxation times of the fast and slow (structural) relaxations, and the microscopic longitudinal kinematic viscosity in Sec. 4. A conclusion is given in Sec. 5.

\section{Generalized Langevin analysis with a sparse modeling}
The experimental data were analyzed by using a GLF \cite{Boon} to clarify several new information about the dynamics of liquid CCl$_4$ as was applied to liquid acetone \cite{HosokawaAcetone}. The formalism starts from intermediate scattering function,
$$F(Q,t)=\frac{1}{N}\sum_{i,j}\langle e^{-\vec{Q}\cdot\vec{r}_i(0)}e^{+\vec{Q}\cdot\vec{r}_j(t)}\rangle,$$
where $\vec{r}_j(t)$ denotes the position of $j$-th particle at a time $t$ in a liquid with $N$ particles with a mass of $m$. In the Langevin formalism, the $F(Q,t)$ function can be calculated by an equation of motion for a damped vibration expressed as,
\begin{equation}
\ddot{F}(Q,t)+\omega_0^2F(Q,t)+\int_0^tM(Q,t-t')\dot{F}(Q,t'){\rm d}t'=0,
\label{generalizedLangevinEquation}
\end{equation}
where $M(Q,t)$ is a memory function of the density fluctuations, and $\omega_0^2=k_BTQ^2/mS(Q)$ is the reduced second frequency moment of $S(Q,\omega)$, giving the lowest limit of sound velocity, i.e., an isothermal sound velocity, at a finite $Q$ as $c_0(Q)=\omega_0(Q)/Q$. The $S(Q,\omega)$ is the frequency spectrum of $F(Q,t)$, and the initial value of $F(Q,t=0)$ corresponds to $S(Q)$. 

Concerning the $M(Q,t)$ function, we used a well-known approximation containing exponential decay channels for a thermal relaxation and two viscoelastic relaxations \cite{Lavesque}, expressed as
\begin{eqnarray}
M(Q,t)&=&[\gamma(Q)-1]\omega_0^2(Q) {\textrm e}^{-\gamma D_T(Q)Q^2t}
\nonumber\\
&+&[\omega_l^2(Q)-\gamma(Q)\omega_0^2(Q)]\nonumber\\
&&\times\{[1-A(Q)]{\textrm e}^{-t/\tau_{\mu}(Q)}
+A(Q){\textrm e}^{-t/\tau_{\alpha}(Q)}\},
\label{MemoryFunction}
\end{eqnarray}
where $\gamma$ is the specific heat ratio at constant pressure and constant volume, and $D_T$ is the thermal diffusivity. In the second term, $\omega_l$ is the reduced forth  moment of $S(Q,\omega)$, which characterizes the prompt response at a finite $Q$, and determines the infinite-frequency sound velocity, $c_l=\omega_l(Q)/Q$. $\tau_\mu$ and $\tau_\alpha$ are respectively the relaxation times for the microscopic $\mu$-relaxation process for a fast relaxation over a very short timescale and for the $\alpha$-relaxation (structural relaxation) process responsible for a long-lasting tail. $A(Q)$ measures the relative weight of the slow decay channel. 

From Eq. (\ref{generalizedLangevinEquation}), a Laplace transform of $F(Q,t)$ is given as \cite{Boon},
$$\tilde{F}(Q,z)=\int_0^\infty{\rm d}t\;e^{-zt}F(Q,t)=S(Q)\left
\{z+\frac{\omega_0^2}{[z+\tilde{M}(Q,z)]}\right\}^{-1}. $$
One can obtain 
\begin{equation}
S(Q,\omega)=\frac{1}{\pi}{\rm Re}\tilde{F}(Q,z=i\omega).
\label{SQw}
\end{equation}
For simplicity, Eq. (\ref{MemoryFunction}) is rewritten as \cite{ScopgnoLiJPCM} 
\begin{equation}
M(Q,t)=\Delta_{th}^2(Q){\textrm e}^{-\gamma D_T(Q)Q^2t}+\Delta_\mu^2(Q){\textrm e}^{-t/\tau_\mu(Q)}+\Delta_\alpha^2(Q){\textrm e}^{-t/\tau_\alpha(Q)}.
\label{DeltaForm}
\end{equation}
By replacing the parameters as in (\ref{DeltaForm}), the $\omega_l$ value can be simply rewritten as, $$\omega_l=\sqrt{\Delta_\mu^2+\Delta_\alpha^2+\gamma\omega_0^2}.$$
Finally, Eq. (\ref{SQw}) can be expressed as \cite{ScopgnoLiJPCM}
\begin{equation}
S(Q,\omega)=\frac{S(Q)}{\pi}{\rm Re}\left\{i\omega+\frac{\omega_0^2}{i\omega+\frac{\Delta_{th}^2}{i\omega+\gamma D_T(Q)Q^2}+\frac{\Delta_\mu^2}{i\omega+1/\tau_\mu}+\frac{\Delta_\alpha^2}{i\omega+1/\tau_\alpha}}\right\}^{-1}
\label{FinalEquation}
\end{equation}
This model function for $S(Q,\omega)$ was convoluted with the Bose factor,
$$\left.\frac{\hbar\omega}{k_BT}\middle/\left(1-e^{-\frac{\hbar\omega}{k_BT}}\right)\right.,$$ 
giving the temperature balance, and the experimentally obtained resolution functions measured by the scattering of a Plexiglas, and then, fitted to the experimental spectra to obtain the dynamical parameters in Eq. (\ref{FinalEquation}).

The $Q$-dependent (generalized) longitudinal kinematic viscosity, $\nu(Q)$, in the GLF analysis corresponds to the total area of the viscoelastic parts of the memory function, which is given by \cite{ScopgnoLiJPCM}
\begin{equation}
\nu(Q)=\frac{\Delta_\mu^2\tau_\mu+\Delta_\alpha^2\tau_\alpha}{Q^2}
\label{KinematicViscosity}
\end{equation}

It should be noted that the GLF was originally considered for {\it monatomic} liquids, and thus, additional contributions of intramolecular interactions to $S(Q,\omega)$ are included for multi-component molecular liquids. Since the intermolecular distance in liquid CCl$_4$ is 0.577 nm \cite{Nishikawa}, however, the intermolecular interactions are dominant in $S(Q,\omega)$ in the low $Q$ region below 11 nm$^{-1}$. Also, the contribution of the localized modes in molecules to $S(Q,\omega)$ is known to be very weak compared with that of LA phonon excitation modes \cite{InuiSe}. 

\begin{figure}
\begin{center}
\includegraphics[width=110mm]{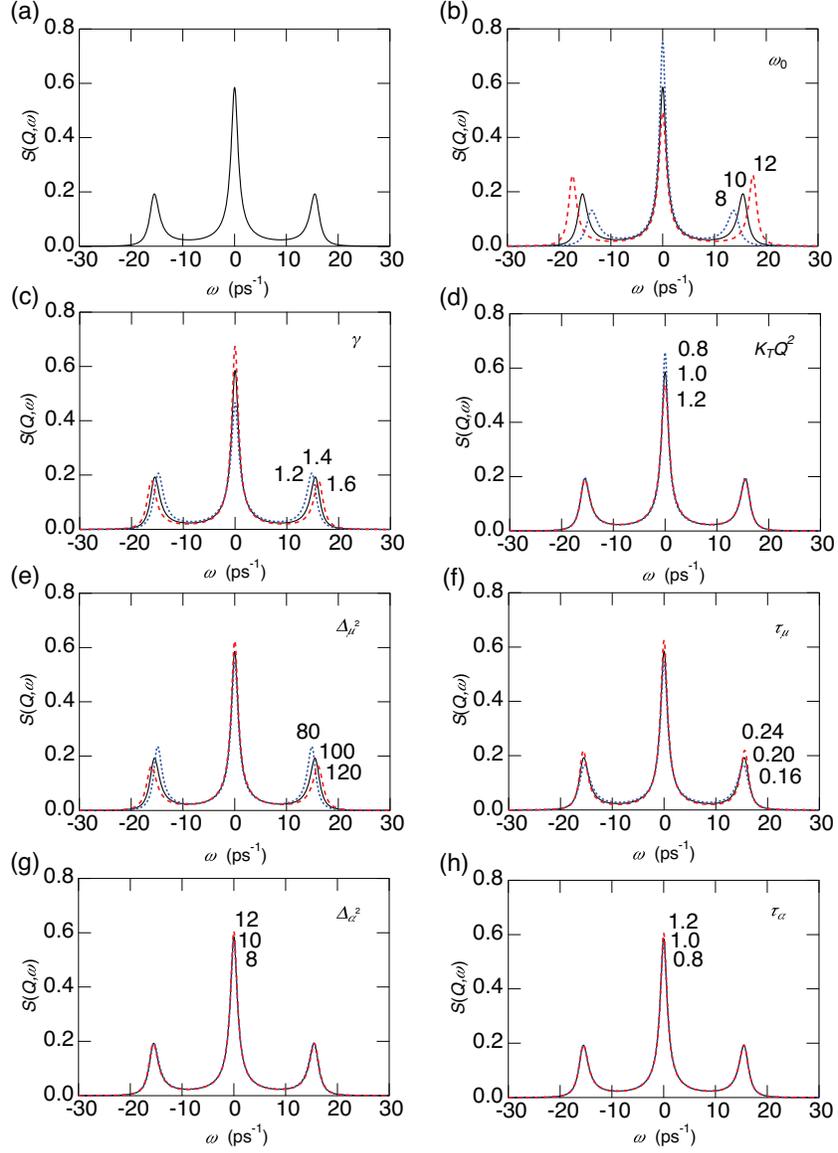}
\caption{\label{gLf_parameter} (a) $S(Q,\omega)$ obtained from the GLF analysis provided $\omega_0=10$ ps$^{-1}$, $\gamma=1.4$, $D_TQ^2=1$ nm$^2$ps$^{-1}$, $\Delta_\mu^2=100$ ps$^{-2}$, $\tau_\mu=0.2$ ps, $\Delta_\alpha^2=10$ ps$^{-2}$, and $\tau_\alpha=1$ ps. The remaining figures show the spectral changes when the parameters vary positively (dashed curves) and negatively (dotted curves) by about 20\% for (b) $\omega_0$, (c) $\gamma$, (d) $D_TQ^2$, (e) $\Delta_\mu^2$, (f) $\tau_\mu$, (g) $\Delta_\alpha^2$, and (h) $\tau_\alpha$, respectively. }
\end{center}
\end{figure}

As seen in Eqs. (\ref{MemoryFunction}) and (\ref{FinalEquation}), there are seven independent dynamical parameters to build the IXS spectral features besides two equipment parameters of the normalization constant $C$ and the energy shift $\Delta\omega$. Here, we examine how the parameters affect the IXS spectrum. We chose $\omega_0$, $\gamma$, $D_TQ^2$, $\Delta_\mu^2$, $\tau_\mu$, $\Delta_\alpha^2$, and $\tau_\alpha$ with $C=1$ and $\Delta\omega=0$. Figure \ref{gLf_parameter}(a) shows an example of $S(Q,\omega)$ given with $\omega_0=10$ ps$^{-1}$, $\gamma=1.4$, $D_TQ^2=1$ nm$^2$ps$^{-1}$, $\Delta_\mu^2=100$ ps$^{-2}$, $\tau_\mu=0.2$ ps, $\Delta_\alpha^2=10$ ps$^{-2}$, and $\tau_\alpha=1$ ps. As seen in the figure, the $S(Q,\omega)$ spectrum is composed of the central quasielastic peak and two inelastic peaks at both the sides of the quasielastic peak. From the peak of the current-current correlation function, $J(Q,\omega)=(\omega/Q)^2S(Q,\omega)$, the excitation energy, $\omega_Q$, is evaluated as the peak position, and the value of 15.57 ps$^{-1}$ (= 10.25 meV) is obtained in the above conditions.

Spectral shapes of the intensities and widths of the quasielastic and inelastic peaks are controlled by the above seven parameters. Figures \ref{gLf_parameter}(b)-(h) show the spectral changes when the parameters vary positively (dashed curves) and negatively (dotted curves) by about 20\% from the original spectra (solid curves) for (b) $\omega_0$, (c) $\gamma$, (d) $D_TQ^2$, (e) $\Delta_\mu^2$, (f) $\tau_\mu$, (g) $\Delta_\alpha^2$, and (h) $\tau_\alpha$. Concerning the quasielastic peak, the height increases with $\gamma$, $\Delta_\mu^2$, and $\tau_\mu$, decreases with $\omega_0$ and $D_TQ^2$, and mostly does not changes with $\Delta_\alpha^2$, and $\tau_\alpha$. Concerning the inelastic excitation mode, the energy position of the excitation increases with $\omega_0$, $\gamma$, $\Delta_\mu^2$, and mostly unchanges with $D_TQ^2$, $\tau_\mu$, $\Delta_\alpha^2$, and $\tau_\alpha$.

It is clearly seen that the spectral shape characteristics always depend on some parameters, and the changes in $\Delta_\alpha^2$, and $\tau_\alpha$ are tiny. Thus, the parameters frequently interfere with each other, which give scattered $Q$ dependences as shown in Ref. \cite{YoshidaBenzene}. Since the fit parameters are assumed to gradually change with $Q$, an idea of sparse modeling can be applied to the data fitting as done on liquid acetone \cite{HosokawaAcetone}. The analytical technique is called `least absolute shrinkage and selection operator (LASSO)' proposed by Tibshirani \cite{Tibshirani}.

In the fitting procedure, the error function is defined as,
\begin{equation}
E=\sum_i|S(Q,\omega_i)-\hat{S}(Q,\omega_i)|^2+\lambda\sum_j|\frac{p_j-p_{j0}}{p_{j0}}|,
\label{SpM}
\end{equation}
where $S(Q,\omega_i)$ and $\tilde{S}(Q,\omega_i)$ are the experimental and model functions with the experimental $\omega_i$ values, $p_j$ are the fit parameters, $p_{j0}$ are the smoothed values of $p_j$ by cubic function fits, and $\lambda$ is a penalty value of the fitting. Only one $\lambda$ value was used for simplicity. When $\lambda=0$, the fit becomes a usual least square fit. We started the results at $\lambda=0$, and the obtained parameters were smoothed over $Q$ by polynomials with a degree of three. Here, parameters were excluded from fitting if the values extremely deviate from reliable values. Then, the fits to the data and optimizations of the $p_{j0}$ values were iterated by increasing $\lambda$ so that $E$ did not exceed the original error by 3\%, corresponding to the error in the experimental data. 

\section{Results}
\begin{figure}[h]
\begin{center}
\includegraphics[width=80mm]{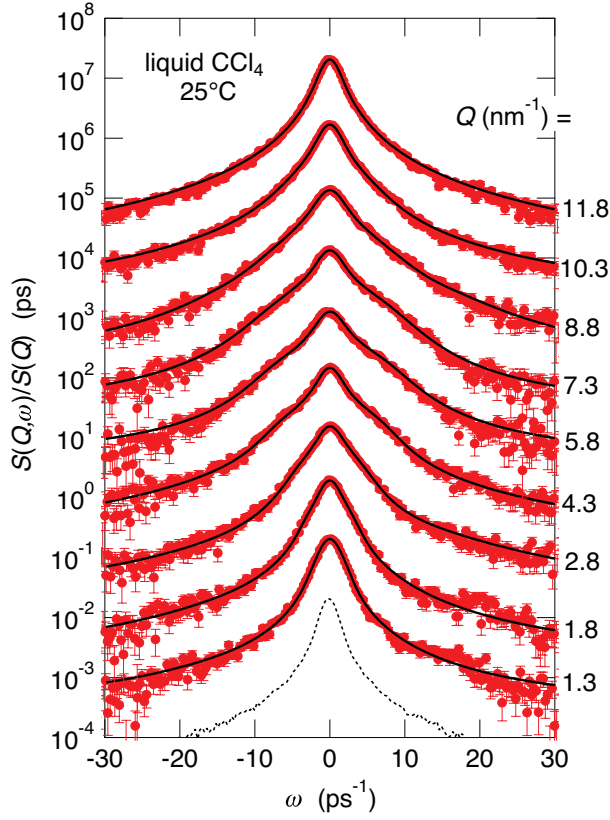}
\caption{\label{SQwexp} Circles show IXS data of liquid CCl$_4$ at selected $Q$ values, previously reported in Ref. \cite{KamiyamaCCl4JPSJ}, and is nearly identical to $S(Q,\omega)/S(Q)$. The solid curves indicate the best fits of the present GLF analysis convoluted with the resolution function given by the dotted curves to the experimental data. For clarity, the data are displaced by one order.}
\end{center}
\end{figure}

Circles in Fig. \ref{SQwexp} show the IXS data of liquid CCl$_4$ at selected $Q$ values normalized with the corresponding $\omega$-integral of $S(Q,\omega)$, which were previously reported in Ref. \cite{KamiyamaCCl4JPSJ}. Since the primary peak in $S(Q)$ is located at about $Q=12$ nm$^{-1}$ \cite{Misawa, Pusztai, Pothoczki}, the $Q$ range in the figure covers the quasi-first and -second Brillouin zones. These curves are nearly identical to $S(Q,\omega)/S(Q)$ except the resolution broadening. The dotted curve is a typical example of the resolution function. Although inelastic excitation modes are not so clear in this figure, small but clear indications of excitation modes are seen at both the sides of the central (quasielastic) peaks, in particular, in the low $Q$ range of 3-8 nm$^{-1}$.

The solid curves in Fig. \ref{SQwexp} indicate the best fits of the present GLF analysis convoluted with the resolution function to the experimental data. Although the GLF analysis was considered as a viscoelastic analysis for simple fluids like liquid metals, the obtained functions fit considerably well to the experimental $S(Q,\omega)$ data of liquid CCl$_4$ over the $Q$ range measured. Recently, applications of GLF have frequently been attempted to macromolecules \cite{MChen}.

\begin{figure}[h]
\begin{center}
\includegraphics[width=120mm]{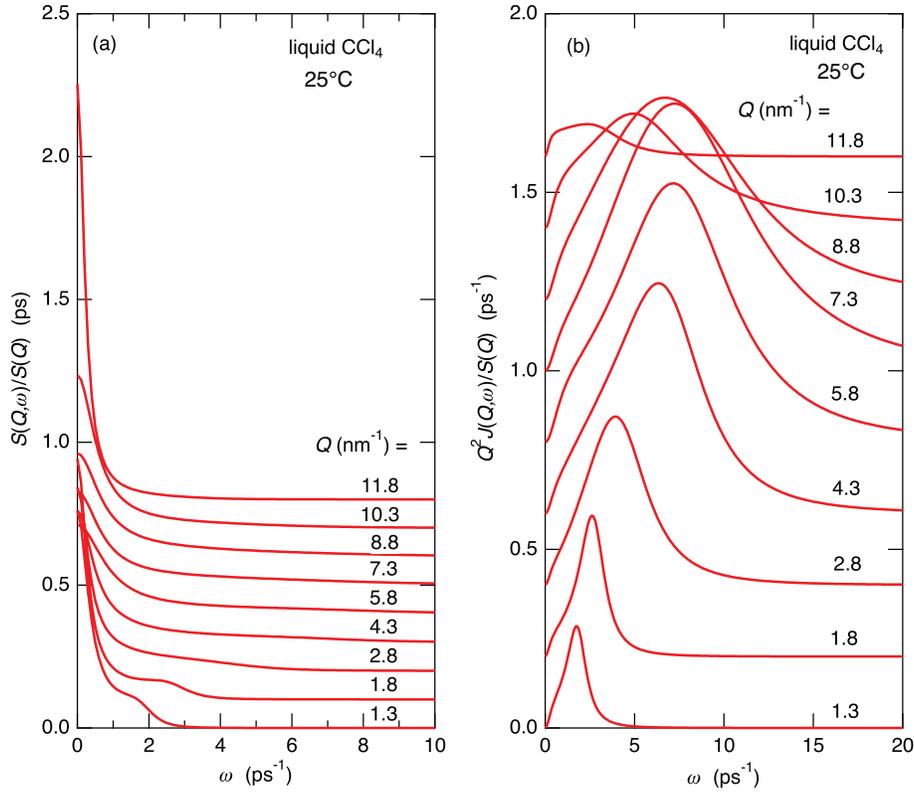}
\caption{\label{SQwJQw} (a) $S(Q,\omega)/S(Q)$ and (b) $Q^2S(Q,\omega)/S(Q)$ at selected $Q$ values obtained from the GLF analysis. For clarity, the data are displaced by (a) 0.1 and (b) 0.2.}
\end{center}
\end{figure}

The existence of LA excitation modes becomes rather clear for the resolution deconvoluted $S(Q,\omega)$ spectra by the GLF analysis as shown in Fig. \ref{SQwJQw}(a) at selected $Q$ values. Although the magnitudes of the inelastic excitations are very small compared with the quasielastic peaks, inelastic shoulders are clearly recognized in the small $Q$ region up to 3 nm$^{-1}$, which show a clear dispersion relation. The shoulder becomes broadened beyond about 3 nm$^{-1}$, being a typical damping with $Q$. 

To obtain the excitation energies of the LA phonons, $\omega_Q$, more clearly, current-current correlation functions, 
$$J(Q,\omega)=\frac{\omega^2}{Q^2}S(Q,\omega),$$ 
were calculated, and the $\omega_Q$ values are defined as the peak energy of the $J(Q,\omega)$ functions. Figure \ref{SQwJQw}(b) shows the $J(Q,\omega)$ functions at the selected $Q$ values multiplied by $Q^2$ and normalized to $S(Q)$ for clarity. As clearly seen in the figure, the peak position, $\omega_Q$ increases with increasing $Q$ up to about 7 nm$^{-1}$, and decreases beyond then. 

\begin{figure}[h]
\begin{center}
\includegraphics[width=60mm]{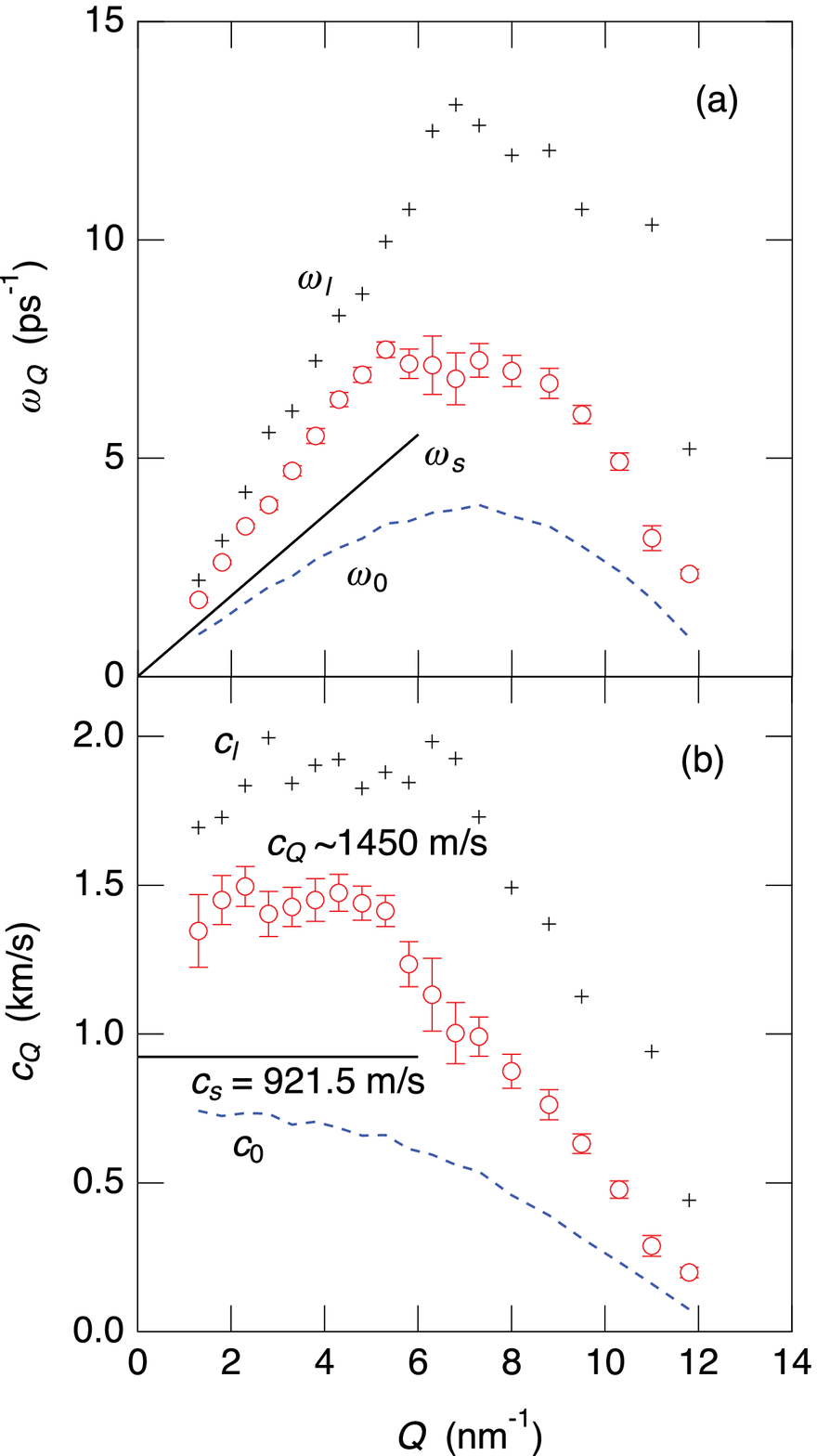}
\caption{\label{wQcQ} (a) $\omega_Q$ and (b) $c_Q$ values obtained from the present GLF analysis. The solid lines indicate the hydrodynamic predictions obtained from the sound velocity \cite{Kolhapurkar}. The dashed lines and crosses represent the corresponding minimum and maximum values, respectively.}
\end{center}
\end{figure}

Figure \ref{wQcQ}(a) shows the $Q$ dependence of $\omega_Q$, i.e. the dispersion relation, given by the open circles. With increasing $Q$, $\omega_Q$ increases in the small range up to about 5 nm$^{-1}$, shows mostly a plateau in the intermediate range of 5-8 nm$^{-1}$, and decreases beyond 8 nm$^{-1}$. Note that the first maximum position in $S(Q)$ is located at about 12 nm$^{-1}$ \cite{Misawa}, and these features in the dispersion relation is typically observed  in liquid materials in the quasi-first and -second Brillouin zones \cite{Boon}. 

The solid line in Fig. \ref{wQcQ}(a) denotes the hydrodynamic prediction of the dispersion relation obtained from a sound velocity result of 921.5 m/s at 25$^\circ$C \cite{Kolhapurkar}. The slope of the line is given by the adiabatic velocity of sound. As clearly seen in the figure, the $\omega_Q$ values in the small $Q$ region remarkably deviate from the hydrodynamic prediction towards the positive direction, which is frequently observed in the terahertz oscillation spectra of liquids, and called {\it positive deviation} or {\it fast sound}. In addition, the lowest and highest limits of the excitation energies, $\omega_0$ and $\omega_l$, obtained from the present GLF analysis are also exhibited by the dashed curve and crosses in the figure, respectively. 

To obviously indicate the information about sound velocity, the dynamical sound velocity is defined as $c_Q=\omega_Q/Q$, and shown in Fig. \ref{wQcQ}(b) by the open circles. With increasing $Q$, the $c_Q$ value seems to slightly increases in the very small region up to 2 nm$^{-1}$, keeps a value of about 1,450 m/s in a wide $Q$ range up to about 5.5 nm$^{-1}$, and then starts to rapidly decrease in the remaining $Q$ domain. The solid line represents the hydrodynamic sound velocity value of $c_0=921.5$ m/s at 25$^\circ$C \cite{Kolhapurkar}. It becomes much prominent that the terahertz $c_Q$ values are much larger than the $c_0$ value by about 57\%. This value is quite lager than our previous result of about 37\% by the DHO analysis \cite{KamiyamaCCl4JPSJ}, although the experimental data are the same. Note that this positive value is similar to that of liquid acetone of about 65\% \cite{HosokawaAcetone} and liquid benzene of about 50\% \cite{YoshidaBenzene} analyzed by the same GLF. It is interesting that molecular liquids generally show large magnitudes of positive dispersions although the intermolecular interactions are quite different from each other. 

In addition, the lowest and highest limits of the sound velocity, $c_0$ and $c_l$, are also depicted in the figure by the dashed curve and crosses, respectively. The $Q\rightarrow0$ limit of $c_0$ is about 750 m/s, slower than the $c_s$ value of 921.5 m/s. The ratio of the adiabatic and isothermal sound velocities is about 1.23, which is caused by the $\gamma$ value as $\sqrt{\gamma}$. Thus, the $\gamma$ value can be estimated to be about 1.51, which is similar to the experimentally obtained macroscopic thermodynamic value of 1.45 \cite{Samios}.

\begin{figure}[h]
\begin{center}
\includegraphics[width=70mm]{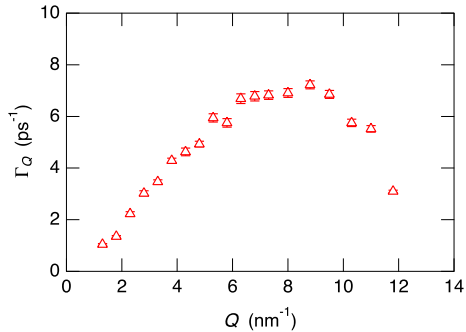}
\caption{\label{GammaQ} The $Q$ dependence of the width of the excitation modes $\Gamma_Q$.}
\end{center}
\end{figure}

Figure \ref{GammaQ} the $Q$ dependence of the width of the LA phonon excitation modes, $\Gamma_Q$, which is defined as the half-width at half-maxima (HWHM) of the $J(Q,\omega)$ functions shown in Fig. \ref{SQwJQw}(b). The $\Gamma_Q$ values are usually proportional to $Q^2$ in the small $Q$ region, and the present results look to follow this relation there. With increasing $Q$, $\Gamma_Q$ has a broad peak at about 8 nm$^{-1}$. The $\Gamma_Q$ value is related to the lifetime of phonons $\tau$ by Heisenberg's uncertainty principle, which will be discussed in the next section.

\section{Discussion}

Scopigno et al. reviewed the experimental results of inelastic experiments on liquid metals and the positive deviations in the range of 15-20\% \cite{ScopignoRev}. An appearance of the positive dispersion for monatomic liquid metals  was interpreted within a framework of generalized hydrodynamic-to-viscoelastic behaviors, where a solid-like shear elasticity should be taken into account on the ps time scale of the terahertz frequency oscillations of atoms \cite{MorkelRbCs}, which is highly related to the existence of transverse excitation modes in liquid metals in the terahertz oscillation region \cite{HosokawaGa, HosokawaSn, HosokawaFeCuZn}.

Molecular liquids generally exhibit much larger effect of the fast sounds, such as water (110\% \cite{SettePRL}), methanol (50\% \cite{YoshidaCPL}), benzene (50\% \cite{YoshidaBenzene}), acetone (57\% \cite{HosokawaAcetone}), normal hexane (75\% \cite{YamaguchiJCP}), cyclohexane (67\% \cite{YamaguchiJCP}), ethylene glycol dimethyl ether (57\% \cite{YamaguchiJCP}), and 1,4-dioxane (65\% \cite{YamaguchiJCP}). Therefore, additional energies are necessary for the terahertz atomic oscillations, which would be highly related to the friction term of the generalized Langevin equation given in Eq. (\ref{generalizedLangevinEquation}). There, the memory function of the frictions includes one thermal and two viscoelastic relaxations. Our previous IXS paper on liquid acetone revealed that the fast and slow viscoelastic relaxation rates of about 0.07 and 0.93 ps match well a vibrational and rotational correlation times of acetone molecules, respectively, and are independent of intermolecular interactions. Thus, it is expected that the extra energies for the terahertz oscillations are consumed through vibrational and rotational movements inside the molecule.  Accordingly, detailed interpretations of the $S(Q,\omega)$ spectra by the GLF analysis are essential to investigate further on the origin of the fast sounds in liquid CCl$_4$. 

\begin{figure}[h]
\begin{center}
\includegraphics[width=65mm]{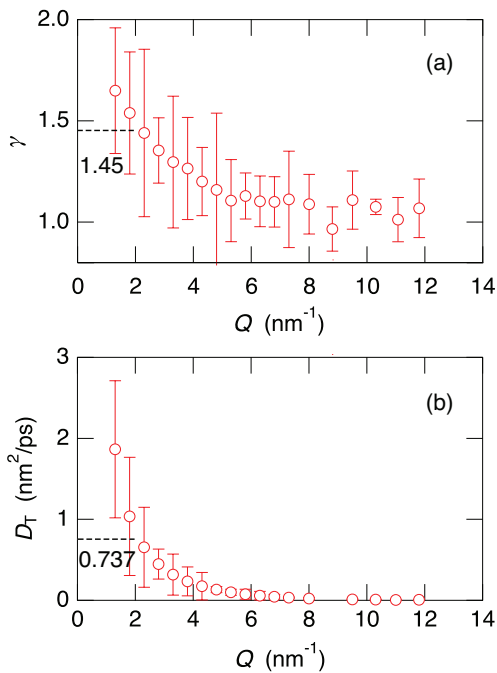}
\caption{\label{gammaDT} $Q$ dependence of (a) $\gamma$ and (b) $D_{\rm T}$ obtained from the present GLF analysis. The dashed lines indicate the hydrodynamic values \cite{Samios}.}
\end{center}
\end{figure}

Figure \ref{gammaDT} shows the $Q$ dependence of the thermal parameters of (a) $\gamma$ and (b) $D_{\rm T}$ obtained from the present GLF analysis. The dashed lines indicate the hydrodynamic values  at $Q\rightarrow0$ \cite{Samios}. Since the contribution of thermal relaxation process to $S(Q,\omega)$ is vary small (in the 1\% range) compared with those of viscoelastic relaxations in GLF analyses \cite{HosokawaFePRB, YoshidaCuNano}, the errors in $\gamma$ and $D_{\rm T}$ are very large. Nevertheless, the microscopic values obtained from the present analysis are consistent with the macroscopic hydrodynamic values indicated by dashed lines. Accordingly, the remaining parameters concerning the viscoelastic relaxations are much reliable. 

\begin{figure}
\begin{center}
\includegraphics[width=65mm]{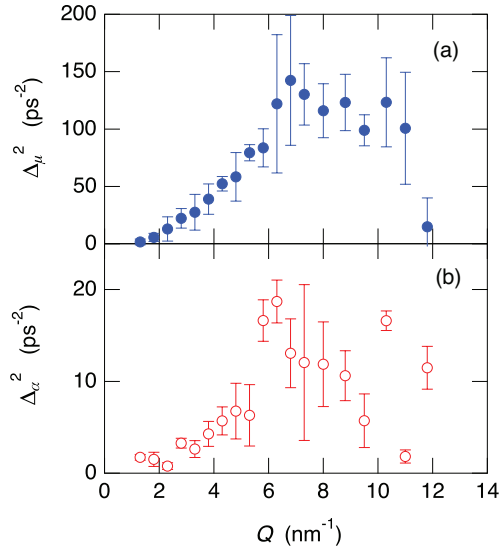}
\caption{\label{W_mualpha} $Q$ dependence of (a) $\Delta_\mu^2$ and (b) $\Delta_\alpha^2$.}
\end{center}
\end{figure}

Figure \ref{W_mualpha} shows the $Q$ dependence of (a) $\Delta_\mu^2$ and (b) $\Delta_\alpha^2$ indicating the initial magnitudes of fast and slow viscoelastic decay channels in the memory function, respectively. As seen in the figures, the $\Delta_\mu^2$ values are about one order of magnitude larger than $\Delta_\alpha^2$. Both the values increase with $Q$ up to about 6-7 nm$^{-1}$ and then decrease up to about 12 nm$^{-1}$, which 
roughly resemble a feature of the dispersion relation given in Fig. \ref{wQcQ}(a). 

\begin{figure}
\begin{center}
\includegraphics[width=65mm]{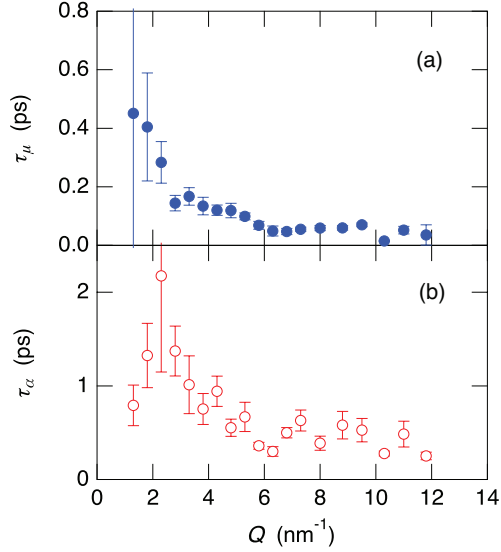}
\caption{\label{tau_mualpha} $Q$ dependence of (a) $\tau_\mu$ and (b) $\tau_\alpha$.}
\end{center}
\end{figure}

Figure \ref{tau_mualpha} shows the $Q$ dependence of (a) $\tau_\mu$ and (b) $\tau_\alpha$. With increasing $Q$, the $\tau_\mu$  value rapidly decreases from about 0.5 ps at the hydrodynamic limit of $Q\sim0$ nm$^{-1}$ to about 0.05 ps at $Q\sim6$ nm$^{-1}$. In case of liquid acetone \cite{HosokawaAcetone}, the $Q\rightarrow0$ limit of $\tau_\mu\sim0.07$ ps corresponds to the stretching vibrational dynamics of the C=O dipoles obtained by a polarized Raman scattering and a MD simulation \cite{Torii}. Raman peaks of liquid CCl$_4$ were compiled by Shimaniuchi \cite{Shimaniuchi} to be 217.0, 313.5, 458.7, and 761.7-790.4 cm$^{-1}$, which were assigned as intramolecular modes of bending, asymmetric stretching, symmetric stretching, and combinations of bending and symmetric stretching, respectively. In addition, a broad low-frequency mode was observed at the peak position of about 25 cm$^{-1}$, which is assigned as the sum of intermolecular modes of oscillational-like and diffusive relaxation-like \cite{Amo}. A oscillational mode was calculated by a  MD simulation and a broad peak was observed at about 40 cm$^{-1}$ \cite{Jo}.
 
Here, we calculate characteristic times of these excitation energies. The intermolecular oscillation with the excitation energy of 40 cm$^{-1}$ corresponds to the time of 0.83 ps, and the intramolecular excitation energies are equivalent to 0.15, 0.11, 0.073, and 0.042-0.044 ps in order of the magnitude of energy. In Fig. \ref{tau_mualpha}(a), the $\tau_\mu$ values at the small $Q$ values of less than 2 nm$^{-1}$ is 0.4-0.5 ps. These values seem to be an average of correlation times of inter- and intramolecular oscillations. The $\tau_\mu$ value rapidly decreases beyond 2.5 nm$^{-1}$ to be the intramolecular bending value of about 0.15 ps. Then, it gradually decreases to be the asymmetric stretching value of 0.11 ps at $Q=5.3$ nm$^{-1}$. Beyond $Q=5.5$ nm$^{-1}$, the $\tau_\mu$ value shows the smaller values of about 0.05 ps. 

Concerning the slow viscoelastic decay channel for liquid acetone \cite{HosokawaAcetone}, it was concluded that the smallest $\tau_\alpha$ value of 0.93 ps corresponds to the reorientation correlation time of acetone molecule of 0.75-0.93 ps observed by NMR measurements, optical measurements of Rayleigh scattering and infrared spectroscopy, and a MD simulation. In the present case of liquid CCl$_4$ at about 25$^\circ$C, experimental values for the rotational relaxation times of 1.7 and 1.8 ps were determined by NMR \cite{O'Reilly} and Raman scattering  \cite{Bartoli} measurements, respectively, which is in excellent agreement with a MD simulation result of 1.7 ps \cite{Chahid}. The value of 1.7-8 ps roughly corresponds to the maximum $\tau_\alpha$ value at about $Q=2.3$ nm$^{-1}$ as shown in Fig. \ref{tau_mualpha}(b). Therefore, it was again confirmed that the slow viscoelastic relaxation time is highly related to the reorientation or rotational correlation time of the molecule. 

\begin{figure}
\begin{center}
\includegraphics[width=65mm]{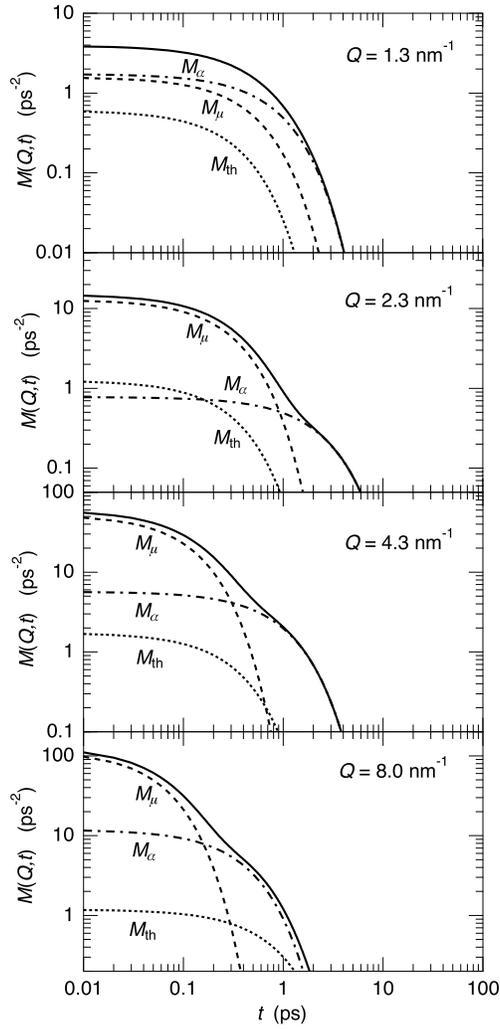}
\caption{\label{CCl4MQt} $M(Q,t)$ functions at selected $Q$ values shown by solid curves. The dotted, dashed, and chain curves indicate the parts of $M_{\rm th}$, $M_\mu$, and $M_\alpha$, respectively.}
\end{center}
\end{figure}

Here, we can discuss the time evolution of the $M(Q,t)$ memory functions obtained from the present GLF analysis as was done in Ref. \cite{HosokawaFePRB}. Solid curves in Fig. \ref{CCl4MQt} shows log-log plots of $M(Q,t)$s at selected $Q$ values of 1.3, 2.3, 4.3, and 8.0 nm$^{-1}$ obtained from the fitting parameters given above by the present GLF analysis. The $M(Q,t)$ functions can be separated into three parts, i.e., the thermal decay channel $M_{\rm th}=\Delta_{\rm th}^2e^{-D_{\rm T}Q^2t}$, the fast viscoelastic one $M_\mu=\Delta_\mu^2e^{-t/\tau_\mu}$, and the slow viscoelastic one $M_\alpha=\Delta_\alpha^2e^{-t/\tau_\alpha}$, which are drawn by dotted, dashed, and chain curves, respectively, in Fig. \ref{CCl4MQt}.

At the very small $Q$ value of 1.3 nm$^{-1}$, $M_\mu$ and $M_\alpha$ compete with each other in the short time range below 0.1 ps, and $M_{\rm th}$ is smaller by a factor of about 5. In the longer time region beyond 0.1 ps, $M_\alpha$ dominates the $M(Q,t)$ function. With increasing $Q$, $M_\mu$ and $M_\alpha$ rapidly and gradually increase at short time below 0.1 ps, while $M_{\rm th}$ remains mostly unchanged in magnitude. In the longer $t$ region, crossovers from $M_\mu$ to $M_\alpha$ are observed. With increasing $Q$, the crossover time decreases from about 2 ps at $Q=2.3$ nm$^{-1}$ to about 0.4 ps at $Q=8.0$ nm$^{-1}$. 

Next, we discuss the $Q$ dependence of the $\tau_\mu$ and $\tau_\alpha$ values shown in Fig. \ref{tau_mualpha}. The $Q$ value is usually considered to correspond to the correlation length $r$ as $2\pi/r$. For the present molecular liquid CCl$_4$, the characteristic $r$ values are the intermolecular C-C distance of 0.54 nm, the intramolecular Cl-Cl distance of 0.289 nm, and the C-Cl nearest neighbor distance of 0.177 nm \cite{Misawa}. The corresponding $Q$ values are estimated to be 11.6, 21.7, and 35.5 nm$^{-1}$ for the C-C, Cl-Cl, and C-Cl distances, which are much larger than the present interest for the $Q$ dependences of the $\tau_\mu$ and $\tau_\alpha$ values.

It is well known that the lifetime, $\tau$, and propagating length, $L$, of phonons in the terahertz oscillation range are very short even in liquid metals \cite{HosokawaGa, HosokawaSn, HosokawaFeCuZn, HosokawaTArev}. The $\tau$ values can be estimated to be, 
$$\tau=\frac{h}{2\hbar\Gamma_Q}=\frac{\pi}{\Gamma_Q}, $$
by taking Heisenberg's uncertainty principle into account, where $h$ and $\hbar$ are Planck's and Dirac's constants ($\hbar=h/2\pi$), respectively, and $\Gamma_Q$ is the HWHM of the phonon excitation peak in $S(Q,\omega)$ shown in Fig. \ref{GammaQ}.  $L$ can be calculated by the product of $\tau$ and the dynamical velocity of sound, $v_Q=\omega_Q/Q$, of the corresponding phonons as
$$L=\tau\cdot\frac{\omega_Q}{Q}=\frac{\pi\omega_Q}{Q\Gamma_Q}.$$
Accordingly, we can figure out which phonons can exist in CCl$_4$ with conditions of $\tau$ and $L$ at a specific $Q$ value. 

\begin{figure}[h]
\begin{center}
\includegraphics[width=65mm]{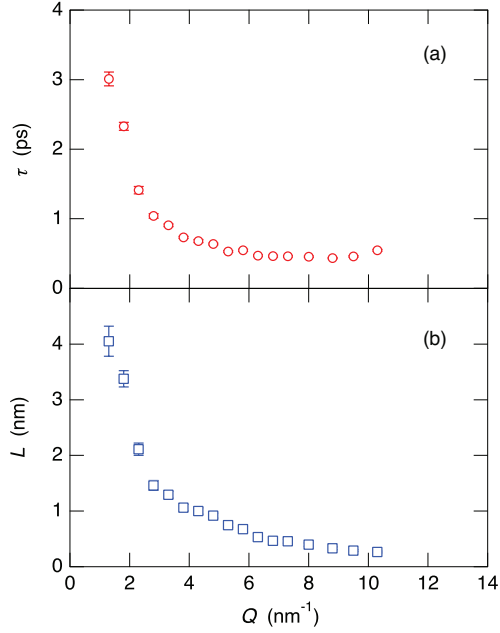}
\caption{\label{tauL_Q} $Q$ dependence of (a) $\tau(Q)$ and (b) $L(Q)$.}
\end{center}
\end{figure}

Figure \ref{tauL_Q} shows the $Q$ dependence of (a) $\tau(Q)$ and (b) $L(Q)$. As clearly seen in the figure, $\tau$ and $L$ are highly $Q$ dependent in the present $Q$ range, which looks similar to the $\tau_\mu$ shown in Fig. \ref{tau_mualpha}(a) having characteristic kink at about 2 and 6 nm$^{-1}$. If $L<0.54$ nm, the excitation mode is limited to a localized one inside the molecules. Accordingly, the excitations beyond about 6 nm$^{-1}$ are highly related to intramolecular localized oscillations, such as symmetric and asymmetric stretching and bending motions inside the CCl$_4$ molecules, whose lifetimes are shorter than 0.5 ps as shown in Fig. \ref{tauL_Q}(a).

In the $Q$ region of 3-6 nm$^{-1}$, the $\tau$ and $L$ values range 0.5-1.0 ps and 0.5-1.5 nm, respectively. In these ranges, phonon oscillations of vibration modes can be coupled with those in the neighboring molecules in addition to the isolated vibrational modes inhabiting in the higher $Q$ range. However, reorientation or rotation motions cannot survive in the allowing $\tau$ range below 1.7-1.8 ps. In the $Q$ region below 2 nm$^{-1}$, the $\tau$ value exceeds 1.5 ps and the $L$ value is beyond 2.0 nm, and the atomic motions include all of the vibrational and rotational modes mentioned above. 

In these interpretations on the viscoelastic decays of the $F(Q,t)$ function by the GLF analysis, it is concluded that the energy losses by the frictions in $F(Q,t)$ are highly related to the vibrational and rotational atomic motions of the CCl$_4$ molecules, which largely enhance the dynamical speed of sound for the terahertz oscillations. Moreover, it can be expected that the major contributions of the fast sound in liquid CCl$_4$ are the vibrational modes coupled with those of neighboring molecules, the reasons of which are as follows. 1) From the comparison of the relaxation rates, it is expected that the vibrational and rotational motions in liquid CCl$_4$ correspond to the $\mu$- and $\alpha$-viscoelastic relaxations in the memory function of the GLF analysis, respectively. 2) As seen in Fig. \ref{wQcQ}(b), the fast sound in the dynamical sound velocity $c_Q$ is observed in the $Q$ range up to 5.5 nm$^{-1}$, where the coupled vibration modes are expected. 3) The large decrease of $c_Q$ beyond $Q\sim5.5$ nm$^{-1}$ is caused by the loss of the coupled vibration contributions in addition to the usual dispersion sinusoidal curve. 4) The rotational atomic motions mainly contribute the $\alpha$-viscoelastic contribution. This contribution is rather small for the fast sound because the $\Delta_\alpha^2$ value only slightly affects the peak position of phonon excitations as seen in Fig. \ref{gLf_parameter}(g), much smaller than that by the $\Delta_\mu^2$ value in Fig. \ref{gLf_parameter}(e). 

\begin{figure}[t]
\begin{center}
\includegraphics[width=65mm]{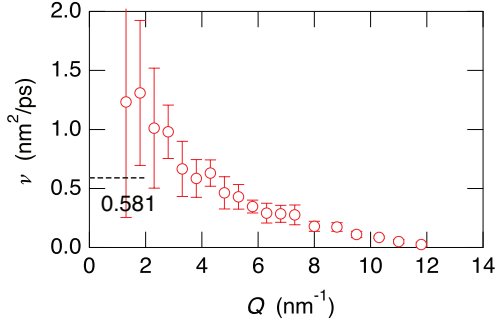}
\caption{\label{nu} $Q$ dependence of $\nu$. The dashed line indicates the shear part of the hydrodynamic value \cite{Samios}.}
\end{center}
\end{figure}

Finally, we show the microscopic longitudinal kinematic viscosity, $\nu(Q)$, calculated by Eq. (\ref{KinematicViscosity}) in Fig. \ref{nu}. The dashed line indicates the shear part of the hydrodynamic value, $4\nu_s/3=0.581$ nm$^2$/ps, indicating the minimum value of the macroscopic kinematic viscosity, where $\nu_s$  is the macroscopic kinematic shear viscosity \cite{Samios}. As usually observed in liquids \cite{HosokawaAcetone, ScopgnoLiJPCM, YoshidaBenzene, HosokawaFePRB}, $\nu$ rapidly decreases with increasing $Q$, and eventually goes to almost zero at the $S(Q)$ maximum. The $Q\rightarrow0$ limit of $\nu$ is about 1.3 nm$^2$/ps, which is reasonable by comparing with the minimum of the hydrodynamic value of 0.581 nm$^2$/ps estimated from the macroscopic shear viscosity \cite{Samios}.

\section{Conclusion}
In this article, we focus on the fast sound observed in IXS spectra in a rather simple molecular liquid of CCl$_4$ by using an improved analytical method of GLF with a memory function containing a thermal and two viscoelastic relaxation channels together with a simple sparse modeling. By using the new method, the obtained excitations of longitudinal acoustic phonons show a largely positive deviation from the hydrodynamic value by about 57\%, larger than about 37\% by the previously reported DHO result. The most important question is why molecular liquids generally exhibit such large values for fast sounds, much larger than 15-20\% of typical liquid metals. The rates of the fast and slow viscoelastic relaxations in the memory function at low $Q$ indicate large values, about 0.5 and 2 ps, which correspond to the vibrational and rotational motions of CCl$_4$ molecules, respectively. Therefore, extra energy-losses for terahertz phonons would be caused by vibrational and rotational motions of molecules. In addition, the $Q$ dependences of the viscoelastic relaxation rates are discussed in terms of lifetime and propagating length of the terahertz oscillations of phonons. Also, the microscopic kinematic longitudinal viscosity rapidly decreases with $Q$ from a reasonable macroscopic value.   

\section*{CRediT authorship contribution statement}
{\bf Shinya Hosokawa} : Conceptualization, Methodology, Formal analysis, Investigation, Data curation, Writing - original draft, Writing - review \& editing, Visualization, Supervision, Project administration, Funding acquisition.  {\bf Koji Yoshida} : Methodology, Formal analysis, Investigation, Data curation, Writing - review \& editing.  

\section*{Declaration of Competing Interest}
The authors declare that they have no known competing financial interests or personal relationships that could have appeared to influence the work reported in this paper.

\section*{Acknowledgments}
SH was supported by JSPS Grant-in-Aid for Transformative Research Areas (A) `Hyper-Ordered Structures Science' (No. 21H05569) and for Scientific Research (C) (No. 22K12662), and by the Japan Science and Technology Agency (JST) CREST (No. JPMJCR1861). KY was supported by JSPS Grant-in-Aid for Scientific Research (C) (Nos. 19K12632 and 22K12673).


\begin{thebibliography}{00}

\bibitem{ArND}
D. G. Henshaw, 
Atomic Distribution in Liquid Argon by Neutron Diffraction and the Cross Sections of A$^{36}$ and A$^{40}$, 
Phys. Rev. {\bf 105}, 976-981 (1957).

\bibitem{Misawa}
M. Misawa, 
Temperature dependence of structure of liquid carbon tetrachloride measured by pulsed neutron total scattering, 
J. Chem. Phys. {\bf 91}, 5648-5654 (1989).

\bibitem{Pusztai}
L. Pusztai and R. L. McGreevy, 
The structure of liquid CCl$_4$, 
Mol. Phys. {\bf 90}, 533-539 (1997).

\bibitem{Pothoczki}
Sz. Pothoczki, L. Temleitner, P. J\'{o}v\"{a}ri, S. Kohara, and L. Pusztai, 
Nanometer range correlations between molecular orientations in liquids of molecules with perfect tetrahedral shape: CCl$_4$, SiCl$_4$, GeCl$_4$,
and SnCl$_4$, 
J. Chem. Phys. {\bf 130}, 064503-1-7 (2009).

\bibitem{Bermejo}
F. J. Bermejo, M. Alvarez, M. Garcia-Hernandez, F. Mompean, R. P. White, W. S. Howells, C. J. Carlile, E. Enciso and F. Batallan, 
Dynamics of liquid CCl$_4$ from quasi-elastic neutron scattering, 
J. Phys.: Condens. Matter {\bf 3}, 851-863 (1991).

\bibitem{Chahid}
A. Chahid, F. J. Bermejo, E. Enciso, M. G. Hernandez and J. L. Martinez, 
Single-particle dynamics of liquid CCl$_4$: a comparison of molecular dynamics and neutron quasi-elastic scattering results, 
J. Phys.: Condens. Matter {\bf 4}, 1213-1231 (1992).

\bibitem{Garcia-Hernandez}
M. Gar\'{c}ia-Hern\'{a}ndez, J. L. Martinez, F. J. Bermejo, A. Chahid and
E. Enciso, 
Collective dynamics of liquid carbon tetrachloride studied by inelastic neutron scattering and computer simulation, 
J. Chem. Phys. {\bf 96}, 8477-8484 (1992).

\bibitem{Burkel}
E. Burkel, 
Phonon spectroscopy by inelastic x-ray scattering, 
Rep. Prog. Phys. {\bf 63}, 171-232 (2000).

\bibitem{KamiyamaCCl4JPSJ}
T. Kamiyama, S. Hosokawa, A. Q. R. Baron, S. Tsutsui, K. Yoshida, W.-C. Pilgrim, Y. Kiyanagi, and T. Yamaguchi, 
Acoustic Phonon Dynamics in Liquid CCl$_4$, 
J. Phys. Soc. Jpn. {\bf 73}, 1615-1618 (2004). 

\bibitem{BaronJPCS}
A. Q. R. Baron, Y. Tanaka, S. Goto, K. Takeshita, T. Matsushita, T. Ishikawa, 
An X-ray scattering beamline for studying dynamics, 
J. Phys. Chem. Solids {\bf 61}, 461-465 (2000).

\bibitem{Fak}
B. F{\aa}k and B. Dorner, 
Phonon line shapes and excitation energies, 
Physica B {\bf 234-236}, 1107-1108 (1997).

\bibitem{ScopignoRev}
T. Scopigno, G. Ruocco, and F. Sette, 
Microscopic dynamics in liquid metals: The experimental point of view, 
Rev. Mod. Phys. {\bf 77}, 881-933 (2005).

\bibitem{MorkelRbCs}
Chr. Morkel, T. Bodensteiner, and H. Gemperlein, 
Zero-sound-like modes in simple liquid metals, 
Phys. Rev. E {\bf 47}, 2575-2580 (1993).

\bibitem{HosokawaTe}
S. Hosokawa, F. Demmel, W.-C. Pilgrim, and F. Albergamo, 
Collective dynamics of liquid Te: The most non-simple liquid metal, 
J. Non-Cryst. Solids {\bf 352}, 5114-5117 (2006).

\bibitem{KajiharaTe}
Y. Kajihara, M. Inui, S. Hosokawa, K. Matsuda, and A. Q. R. Baron, 
Dynamical inhomogeneity of liquid Te near the melting temperature proved by the inelastic x-ray scattering measurements, 
J. Phys.: Condens. Matter {\bf 20}, 494244-1-7 (2008).

\bibitem{HosokawaAcetone}
S. Hosokawa, T. Kamiyama, K. Yoshida, T. Yamaguchi, S. Tsutsui, and A. Q. R. Baron, 
Collective dynamics of liquid acetone investigated by inelastic X-ray scattering, 
J. Mol. Liq. {\bf 332}, 115825-1-9 (2021).

\bibitem{Boon}
J. P. Boon and S. Yip, 
Molecular Hydrodynamics, 
McGraw-Hill, New York, 1980.

\bibitem{Lavesque} 
D. Lavesque, L. Verlet, and J. K\"{u}rkijarvi, 
Computer ``Experiments" on Classical Fluids. IV. Transport Properties and Time-Correlation Functions of the Lennard-Jones Liquid near Its Triple Point, 
Phys. Rev. A {\bf 7}, 1690-1700 (1973).

\bibitem{ScopgnoLiJPCM}
T. Scopigno, U. Balucani, G. Ruocco, and F. Sette, 
Density fluctuations in molten lithium: inelastic x-ray scattering study, 
J. Phys.: Condens. Matter {\bf 12}, 8009-8034 (2000).

\bibitem{Nishikawa}
K. Nishikawa, K. Tohji, M. Shima, and Y. Murata, 
The temperature dependence of the liquid structure of carbon tetrachloride, 
Chem Phys, Lett. {\bf 64}, 154-157 (1979).

\bibitem{InuiSe} 
M. Inui, S. Hosokawa, K. Matsuda, S. Tsutsui, and A. Q. R. Baron, 
Heavy particle dynamics in liquid Se: Inelastic x-ray scattering, 
J. Phys. Soc. Jpn. {\bf 76}, 053601-1-4 (2007).

\bibitem{YoshidaBenzene}
K. Yoshida, N. Fukuyama, T. Yamaguchi, S. Hosokawa, H. Uchiyama, S. Tsutsui, and A. Q. R. Baron, 
Inelastic X-ray scattering on liquid benzene analyzed using a generalized Langevin equation, 
Chem. Phys. Lett. {\bf 680}, 1-5 (2017).

\bibitem{Tibshirani}
R. Tibshirani, 
Regression shrinkage and selection via the lasso, 
J. Royal Stat. Soc. B {\bf 58}, 267-288  (1996).

\bibitem{MChen}
M. Chen, X. Li, and C. Liu, 
Computation of the memory functions in the generalized Langevin models for collective dynamics of macromolecules, 
J. Chem. Phys. {\bf 141}, 064112-1-10 (2014).

\bibitem{Kolhapurkar}
R. R. Kolhapurkar, D. H. Dagade, R. B. Pawar, K. J. Patil, 
Compressibility studies of aqueous and CCl$_4$ solutions of 18-crown-6 at $T$ = 298.15 K, 
J. Chem. Thermodynamics {\bf 38}, 105-112 (2006).

\bibitem{Samios}
D. Samios and Th. Dorfm\"{u}ller, 
A light scattering study of vibrational relaxation in liquids I. Pure tetrachlorides, 
Mol. Phys. {\bf 41}, 637-651 (1980). 

\bibitem{HosokawaGa}
S. Hosokawa, M. Inui, Y. Kajihara, K. Matsuda, T. Ichitsubo, W.-C. Pilgrim, H. Sinn, L. E. Gonz\'{a}lez, D. J. Gonz\'{a}lez, S. Tsutsui, and A. Q. R. Baron, 
Transverse Acoustic Excitations in Liquid Ga, 
Phys. Rev. Lett. {\bf 102}, 105502-1-4 (2009).

\bibitem{HosokawaSn}
S. Hosokawa, S. Munejiri, M. Inui, Y. Kajihara, W.-C. Pilgrim, Y. Ohmasa, S. Tsutsui, A. Q. R. Baron, F. Shimojo, and K Hoshino, 
Transverse excitations in liquid Sn, 
J. Phys.: Condens. Matter {\bf 25}, 112101-1-5 (2013). 

\bibitem{HosokawaFeCuZn}
S. Hosokawa, M. Inui, Y. Kajihara, S. Tsutsui, and A. Q. R. Baron, 
Transverse excitations in liquid Fe, Cu and Zn, 
J. Phys.: Condens. Matter {\bf 27}, 194104-1-7 (2015).

\bibitem{SettePRL}
F. Sette, G. Ruocco, M. Krisch, C. Masciovecchio, R. Verbeni, and U. Bergmann, 
Transition from $Normal$ to $Fast$ Sound in Liquid Water, 
Phys. Rev. Lett. {\bf77} 83-86 (1996). 

\bibitem{YoshidaCPL}
K. Yoshida, N. Yamamoto, S. Hosokawa, A. Q. R. Baron, and T. Yamaguchi, 
Collective dynamics of sub- and supercritical methanol by inelastic X-ray scattering, 
Chem. Phys. Lett. {\bf 440}, 210-214 (2007).

\bibitem{YamaguchiJCP}
Ts. Yamaguchi, K. Yoshida, S. Hosokawa, D. Ishikawa, and A. Q. R. Baron, 
Effects of molecular shape and flexibility on fast sound of organic liquids, 
J. Chem Phys. {\bf 157}, 154504-1-11 (2022). 

\bibitem{HosokawaFePRB}
S. Hosokawa, M. Inui, K. Matsuda, D. Ishikawa, and A. Q. R. Baron, 
Damping of the collective modes in liquid Fe, 
Phys. Rev. B {\bf77}, 174203-1-10 (2008).

\bibitem{YoshidaCuNano}
K. Yoshida and T. Yamaguchi, 
Generalized Langevin analysis of inelastic X-ray scattering for copper/ethylene glycol nanofluid, 
Chem. Phys. Lett. {\bf 718}, 74-79 (2019).

\bibitem{Torii}
H. Torii, M. Musso, and M. G. Giorgini, 
Modulations of vibrational frequencies and other vibrational properties of the C=O stretching mode of liquid acetone, 
J. Mol. Liq. {\bf134}, 129-135 (2007).

\bibitem{Shimaniuchi}
T. Shimaniuchi, 
Tables of Molecular Vibrational Frequencies Consolidated Volume I, National Bureau of Standards, Gaithersburg, 1972, pp. 1-160.

\bibitem{Amo}
Y. Amo and Y. Tominaga, 
Low-frequency Raman scattering of liquid CCl$_4$, CHCl$_3$, and acetone, 
J. Chem. Phys. {\bf 109}, 3994-3998 (1998).

\bibitem{Jo}
J.-Y. Jo, H. Ito, and Y. Tanimura, 
Full molecular dynamics simulations of liquid water and carbon tetrachloride for two-dimensional Raman spectroscopy in the frequency domain, 
Chem Phys. {\bf 481}, 245-249 (2016).

\bibitem{O'Reilly}
D. E. O'Reilly and G. E. Schacher, 
Rotational Correlation Times for Quadrupolar Relaxation in Liquids, 
J. Chem. Phys. {\bf 39}, 1768-1771 (1963). 

\bibitem{Bartoli}
F. J. Bartoli and T. A. Litovitz, 
Raman Scattering: Orientational Motions in Liquids, 
J. Chem. Phys. {\bf 56}, 413-425 (1971). 

\bibitem{HosokawaTArev}
S. Hosokawa, 
Transverse acoustic phonon excitations in liquid metals, 
Z. Phys. Chem. {\bf 235}, 99-115 (2021). 

\end{thebibliography}
\end{document}